\documentclass[twoside,twocolumn,9pt]{article}
\usepackage{extsizes}
\usepackage[super,sort&compress,comma]{natbib} 
\usepackage[version=3]{mhchem}
\usepackage[left=1.5cm, right=1.5cm, top=1.785cm, bottom=2.0cm]{geometry}
\usepackage{balance}
\usepackage{mathptmx}
\usepackage{sectsty}
\usepackage{graphicx} 
\usepackage{lastpage}
\usepackage[format=plain,justification=justified,singlelinecheck=false,font={stretch=1.125,small,sf},labelfont=bf,labelsep=space]{caption}
\usepackage{float}
\usepackage{fancyhdr}
\usepackage{fnpos}
\usepackage[english]{babel}
\addto{\captionsenglish}{%
  
}
\usepackage{array}
\usepackage{droidsans}
\usepackage{charter}
\usepackage[T1]{fontenc}
\usepackage[usenames,dvipsnames]{xcolor}
\usepackage{setspace}
\usepackage[compact]{titlesec}
\usepackage{amssymb}
\usepackage{hyperref}
\usepackage{epstopdf}

\definecolor{cream}{RGB}{222,217,201}

\begin{document}
\pagestyle{fancy}
\thispagestyle{plain}
\fancypagestyle{plain}{

\renewcommand{\headrulewidth}{0pt}
}

\makeFNbottom
\makeatletter
\renewcommand\LARGE{\@setfontsize\LARGE{15pt}{17}}
\renewcommand\Large{\@setfontsize\Large{12pt}{14}}
\renewcommand\large{\@setfontsize\large{10pt}{12}}
\renewcommand\footnotesize{\@setfontsize\footnotesize{7pt}{10}}
\makeatother

\renewcommand{\thefootnote}{\fnsymbol{footnote}}
\renewcommand\footnoterule{\vspace*{1pt}%
\color{cream}\hrule width 3.5in height 0.4pt \color{black}\vspace*{5pt}} 
\setcounter{secnumdepth}{5}

\makeatletter 
\renewcommand\@biblabel[1]{#1}            
\renewcommand\@makefntext[1]%
{\noindent\makebox[0pt][r]{\@thefnmark\,}#1}
\makeatother 
\renewcommand{\figurename}{\small{Fig.}~}
\sectionfont{\sffamily\Large}
\subsectionfont{\normalsize}
\subsubsectionfont{\bf}
\setstretch{1.125} 
\setlength{\skip\footins}{0.8cm}
\setlength{\footnotesep}{0.25cm}
\setlength{\jot}{10pt}
\titlespacing*{\section}{0pt}{4pt}{4pt}
\titlespacing*{\subsection}{0pt}{15pt}{1pt}

\fancyfoot{}

\fancyfoot[RO]{\footnotesize{\sffamily{1--\pageref{LastPage} ~\textbar  \hspace{2pt}\thepage}}}
\fancyfoot[LE]{\footnotesize{\sffamily{\thepage~\textbar\hspace{3.45cm} 1--\pageref{LastPage}}}}
\fancyhead{}
\renewcommand{\headrulewidth}{0pt} 
\renewcommand{\footrulewidth}{0pt}
\setlength{\arrayrulewidth}{1pt}
\setlength{\columnsep}{6.5mm}
\setlength\bibsep{1pt}

\makeatletter 
\newlength{\figrulesep} 
\setlength{\figrulesep}{0.5\textfloatsep} 

\newcommand{\topfigrule}{\vspace*{-1pt}%
\noindent{\color{cream}\rule[-\figrulesep]{\columnwidth}{1.5pt}} }

\newcommand{\botfigrule}{\vspace*{-2pt}%
\noindent{\color{cream}\rule[\figrulesep]{\columnwidth}{1.5pt}} }

\newcommand{\dblfigrule}{\vspace*{-1pt}%
\noindent{\color{cream}\rule[-\figrulesep]{\textwidth}{1.5pt}} }

\makeatother

\noindent\LARGE{\textbf{SrMn$_2$Ge$_2$: A metallic system with anomalous magnetotransport properties}} \\

\vspace{0.3cm}

\noindent\large{A. van der Spuy,$^{\dag}$\textit{$^{a}$} R. Warmbier,\textit{$^{a}$} A. M. Strydom,\textit{$^{b}$} Daniel Wamwangi,\textit{$^{a}$} and Abhishek Pandey,*\textit{$^{a}$}}. \\

\noindent\normalsize {The structural, electronic, and physical properties of previously unexplored layered tetragonal compound SrMn$_2$Ge$_2$ are reported. The compound crystallizes in the ThCr$_2$Si$_2$-type structure with a crystallographic parameters $c/a$ ratio of $2.535(3)$, significantly smaller than those of the related BaMn$_2$$Pn_2$ compounds ($Pn =$ P, As, Sb, Bi), indicating  enhanced interlayer interaction between the moment-bearing Mn-ions. Magnetic susceptibility measurements suggest an antiferromagnetic ground state with a high Ne\'{e}l temperature $T_{\rm N} > 350$~K\@. A large and positive magnetoresistance (MR) of approximately $50\%$ is observed at low temperatures, which follows a power-law field dependence with exponent $n = 1.50(8)$. Low-temperature electrical resistivity data reveal Kondo-like features. Heat capacity measurements show no signs of a phase transition below 300~K and further confirm a metallic ground state. Electronic structure calculations reveal a three-dimensional Fermi surface and suggest the presence of substantial electronic correlation effects.} \\



\renewcommand*\rmdefault{bch}\normalfont\upshape
\rmfamily
\section*{}
\vspace{-1cm}

\footnotetext{\textit{$^{a}$~Materials Physics Research Institute, School of Physics, University of the Witwatersrand, Johannesburg 2000, Gauteng, South Africa.}}
\footnotetext{\textit{$^{b}$Highly Correlated Matter Research Group, Physics Department, University of Johannesburg, Johannesburg 2006, South Africa}}
\footnotetext{\textit{$^{\dag}$E-mail: arthurvanderspuy@gmail.com}}
\footnotetext{\textit{$^{*}$E-mail: abhishek.pandey@wits.ac.za}}

\section{Introduction}

The discovery of high-$T_{\rm c}$ superconductivity \cite{Rotter-2008, Paglione-2010, Johnston-2010,Stewart-2011} and several other exciting ground states, such as highly frustrated magnetic state,\cite{Sapkota-2011} prototypical half-metallic behavior,\cite{Pandey-2013a, Ueland-2015, Pandey-2015} Anderson localization, and large negative magnetoresistance (MR)\cite{Huynh-2019, Ogasawara-2021, Ogasawara-2022} have sparked intense interest in ThCr$_2$Si$_2$-type 122 compounds. Among these compounds, the $R$Mn$_2$Ge$_2$ ($R$: rare earth) series has been shown to exhibit a wide variety of complex magnetic structures and behaviours. \cite{Kolmakova-2002, Duraj-1988, Dover-1993, McLanahan-2025, Welter-1995,Gong-2021,Roychowdhury-2024} These include reentrant ferromagnetism, a variety of metamagnetic transitions, as well as a host of noncollinear structures. \cite{Fujii 1985,Shigeoka 1985, Wang 2022} Furthermore, the subtle and layered nature of the underlying magnetic sublattices in these systems can lead to unusual magnetotransport behaviour. \cite{Mallik 1997, Sampathkumaran 1996} For example, LaMn$_2$Ge$_2$ which exhibits a non-trivial spin structure, has been reported to show anomalously large and positive MR of $\sim80\%$ at 4.2~K in an applied field of $\mu_0H = 5$~T. \cite{Mallik 1997} In addition, this compound has recently been reported to exhibit a large topological Hall effect near room temperature that highlights its technological relevance.\cite{Gong-2021, Roychowdhury-2024} 

The related $A$Mn$_2$Ge$_2$ ($A =$~Ca, Sr, Ba) series remains largely unexplored, with only a few studies on crystal and magnetic structures having been reported so far. Both CaMn$_2$Ge$_2$ and BaMn$_2$Ge$_2$ have been shown to be antiferromagnetic (AFM) with high N\'{e}el temperatures and collinear G-type magnetic ordering, with $T_{\rm N} = 670$~K reported for the former. \cite{Malaman-1994, Welter-1998} On the other hand, only structural properties of SrMn$_2$Ge$_2$ have been reported so far. \cite{Dorrscheidt-1976} Interestingly, all three members of this series feature small $c/a$ ratios, as well as short interlayer Ge-Ge distances $d_{\rm Ge-Ge}$, where the latter is close to the single covalent Ge--Ge bond distance of $\sim 2.45$~\AA\ (Table~\ref{Table:Structure}). These observations collectively suggest the presence of interlayer Ge--Ge dimerization, leading to a collapsed tetragonal (cT) structure in all three $A$Mn$_2$Ge$_2$ compounds, as well as in LaMn$_2$Ge$_2$ and EuMn$_2$Ge$_2$. Furthermore, the $c/a$ ratios of $A$Mn$_2$Ge$_2$ compounds are significantly smaller than those reported for BaMn$_2$$Pn_2$ ($Pn = $ P, As, Sb, Bi), where the magnetism of the Mn-sublattice has been reported to be quasi-two-dimensional in nature. \cite{Jacobs-2023} These crystallographic observations therefore suggest that the underlying  electronic and magnetic interactions in the discussed Ge-compounds are likely to be more three-dimensional (3D) in nature. 

To gain a better understanding of the underlying interactions and properties of Ge-based ThCr$_2$Si$_2$-type 122 compounds, we have investigated previously unexplored compound SrMn$_2$Ge$_2$. Here we report a comprehensive investigation of the physical properties of polycrystalline SrMn$_2$Ge$_2$ using x-ray diffraction, magnetic, thermal, electronic and magneto-transport measurements. Additionally, we also present Density Functional Theory (DFT) based electronic structure calculations and discuss these results in relation to the experimentally obtained data. Our results indicate the presence of an AFM ground state with $T_{\rm N} > 350$~K and show an unusual positive MR of about $50\%$ at low temperatures. The observed unusual MR indicates the presence of correlated multiband electrical transport and/or complex magnetic scatterings which may, in some sense, be similar to those in LaMn$_2$Ge$_2$. Hence, our results suggest that SrMn$_2$Ge$_2$ could be a promising candidate for investigation of topologically relevant electrical transport behavior.  

\section{Experimental and computational details}

Single crystal growth of SrMn$_2$Ge$_2$ is nontrivial. Several attempts using different fluxes, including Sn, In, Ge, and self flux MnGe, were unsuccessful. Similarly, obtaining a high-purity polycrystalline sample of this compound proved highly challenging. Polycrystalline samples of SrMn$_2$Ge$_2$ were synthesized by the arc-melting technique under a partial argon pressure using high-purity constituent elements: Sr (99.9\% from Merck), Mn (99.99\% from Alfa Aesar), and Ge (99.999\% from Alfa Aesar). While Mn and Ge were taken in the stoichiometric ratio, Sr was taken with a varying amount of excess (<7~at.\%) to compensate for the anticipated losses due to evaporation, given its relatively low boiling temperature. Numerous synthesis attempts employing different excess Sr concentrations were performed, and the highest purity sample used in the present study with about approximately 5~wt.\% of Mn$_5$Ge$_3$ was obtained in one of the attempts using ~3~at.\% excess of Sr. The occurrence of the impurity phase Mn$_5$Ge$_3$ was also previously reported in CaMn$_2$Ge$_2$ and BaMn$_2$Ge$_2$ compounds investigated using neutron diffraction measurements.\cite{Malaman-1994}

Room temperature powder x-ray diffraction (XRD) measurements were performed using a Rigaku Miniflex powder diffractometer equipped with a Cu-K$_\alpha$ radiation source. Rietveld refinements were performed using the FullPROF package \cite{Carvajal-1993}.Temperature- and magnetic-field-dependent electrical resistivity and magnetoresistance measurements were performed using the conventional four-probe method in a Physical Properties Measurement System (PPMS) of Quantum Design Inc., USA. Two-component Epo-Tek epoxy was used for connecting the annealed Pt leads to the sample. For field-dependent transport measurements, the magnetic field was applied transverse to the direction of the electrical current. Heat capacity, magnetic susceptibility, and isothermal magnetization measurements were also carried out using the PPMS. For the heat capacity measurements, the sample was thermally anchored to the measurement platform using Apiezon N grease.  The magnetic measurements were performed using the quartz sample holder of the Vibrating Sample Magnetometer Option of the same PPMS.

DFT-based calculations were performed using the GPAW software \cite{Mortensen 2024,Larsen 2017,Enkovaara 2010,Mortensen 2005}. These calculations were performed with r\textsuperscript{2}SCAN meta-GGA exchange-correlation functional \cite{Furness 2020,Furness2 2020} as implemented in libxc \cite{Lehtola 2018}. r\textsuperscript{2}SCAN is an all-purpose meta-GGA functional that often provides better consistency than GGAs for correlated materials. Additional results obtained from calculations using the PBEsol exchange-correlation functional \cite{Perdew 2008,Constantin 2009} are provided as these were found to provide similar agreement with experimental results. Total-energy calculations were performed for the ferromagnetic (FM) configuration, as well as four AFM configurations: A-type, C-type, S-type, and G-type. Among these, the G-type AFM configuration, which is also adopted by CaMn$_2$Ge$_2$ and BaMn$_2$Ge$_2$,\cite{Malaman-1994} yielded the lowest total energy. Accordingly, all subsequent calculations were performed assuming the G-type AFM ground state, using the experimentally determined lattice parameters and atomic positions as input.
Calculations were performed with a plane-wave energy cutoff of 900~eV, as well as $30\times30\times12$ Monkhorst-Pack mesh. A denser $k$-space sampling of 30 points per \AA$^{-1}$ was used for the Density of States (DOS) calculation leading to a spacing of 10~meV per bin for the DOS as a function of energy. Tetrahedron integration, as implemented in GPAW, was used. 

\section{Results}

\subsection{Crystal structure}

\begin{figure}[ht]
\centering
\includegraphics[width=3.0in]{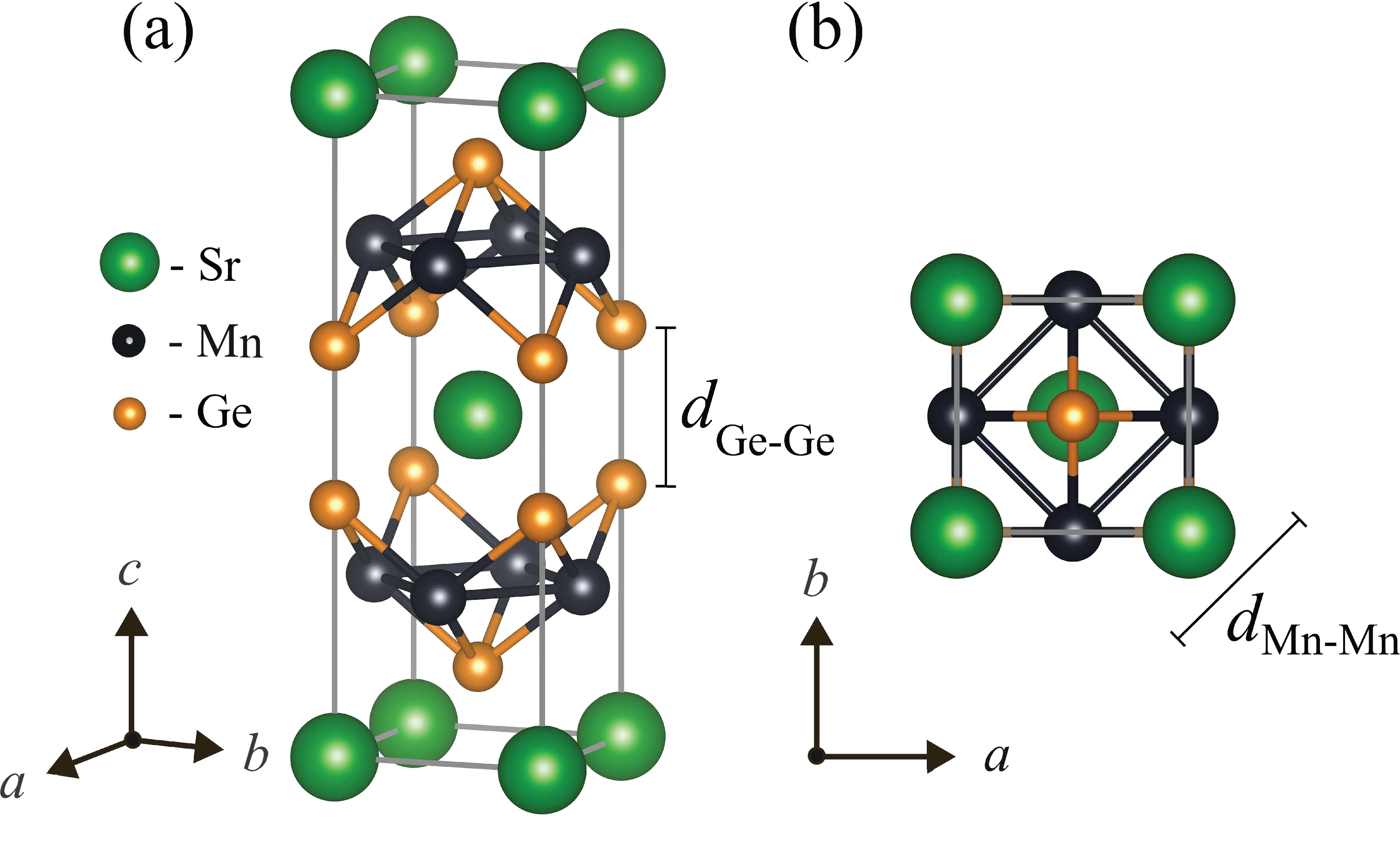}
\includegraphics[width=3.0in]{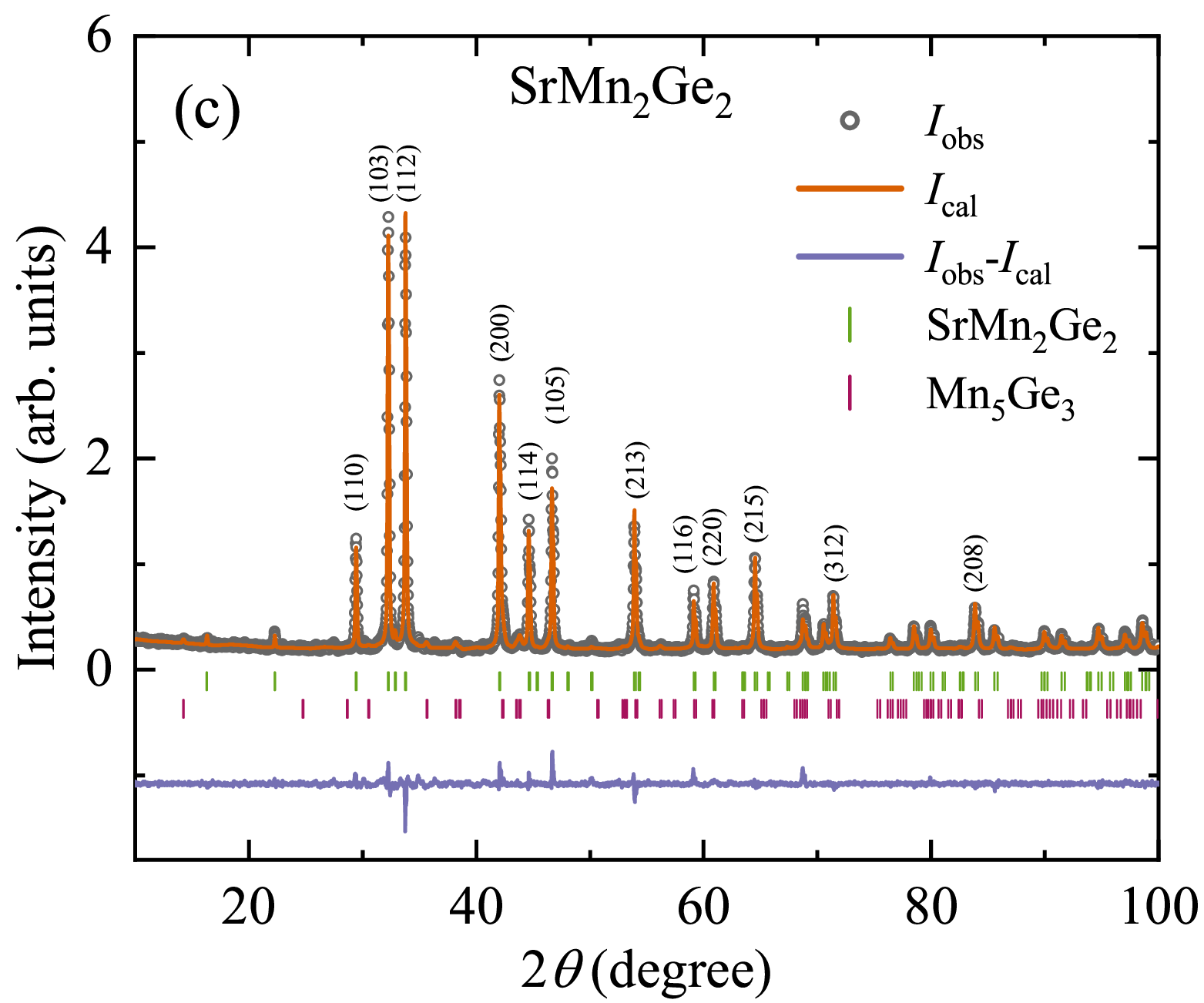}
\caption{(a) ThCr$_2$Si$_2$-type (space group: $I4/mmm$) crystallographic unit cell of SrMn$_2$Ge$_2$. The interlayer Ge-Ge distance is shown. (b) The top view of the unitcell. The intralayer Mn-Mn distance is indicated. (c) Room temperature powder x-ray diffraction data  $I_{\rm obs}$ of SrMn$_2$Ge$_2$, together with two-phase Rietveld refinement fit $I_{\rm cal}$, difference profile $I_{\rm obs}-I_{\rm cal}$ and calculated Bragg peak positions for SrMn$_2$Ge$_2$ (green bars) and impurity phase Mn$_5$Ge$_3$ (purple bars).}
\label{Fig:XRD_Ge}
\end{figure}

\begin{table*} 
\centering 
\caption{Structural parameters of SrMn$_2$Ge$_2$, as well as those reported for $A$Mn$_2$Ge$_2$ ($A = $ Ca, Ba) and $R$Mn$_2$Ge$_2$ ($R = $~La, Eu). All these compounds crystallize with ThCr$_2$Si$_2$-type body centre tetragonal structure [space group: $I4/mmm$ (139)]. The crystallographic parameters provided are the unit cell parameters $a$, $c$, $c/a$, $z_{\rm Ge}$, the intralayer distance between two nearest Mn atoms $d^{ab}_{\rm Mn-Mn} = \frac{1}{\sqrt{2}}a$, and interlayer distance between two nearest Ge atoms $d_{\rm Ge-Ge} = c(1-2z_{\rm Ge})$. }

\label{Table:Structure}
\hspace*{-0.4cm}
\begin{tabular}{l l l l l l }
\hline
     Parameter & CaMn$_2$Ge$_2$ & SrMn$_2$Ge$_2$ & BaMn$_2$Ge$_2$ & LaMn$_2$Ge$_2$  & EuMn$_2$Ge$_2$ \\
		           & Ref.~[\cite{Malaman-1994}] & (This work) & Ref.~[\cite{Malaman-1994}] & Ref.~[\cite{Venturini 1994}] & Ref.~[\cite{Hofmann-2004,Felner-1978}]\\
	 \hline
      $a$ (\AA) &4.153(2) & 4.3052(2) & 4.473(5) & 4.204(2) & 4.24972(3) \\
			$c$ (\AA) & 10.864(4) & 10.9289(4) & 10.98(3) & 10.996(6) & 10.9084(2))\\
			$c/a$ & 2.616(3) & 2.535(3) & 2.455(6) & 2.615(3) & 2.5668(2) \\
			$z_{\rm Ge}$ & 0.379(2) & 0.3798(1) & 0.3778(8) & 0.380(1) & 0.38 \\
			$d^{ab}_{\rm Mn-Mn}$ (\AA) & 2.937(2) & 3.0442(2) & 3.163(4) & 2.973(2) & 3.0050(3) \\
			$d_{\rm Ge-Ge}$ (\AA) & 2.63(4) & 2.627(3) & 2.68(2) & 2.64(2) & 2.618(4) \\
  \hline
            
\end{tabular}
\end{table*}

Rietveld refinement of the room temperature powder XRD of SrMn$_2$Ge$_2$ using the ThCr$_2$Si$_2$-type structure with space group $I4/mmm$ (SG:~139) led to a satisfactory fit to the data [Fig.~\ref{Fig:XRD_Ge}]. However, this refinement revealed a few small impurity peaks in the range $20^{\circ}<2\theta<60^{\circ}$. The most intense impurity peak, with intensity $<2\%$ of the highest intensity peak of the pattern, occurs at $\sim42^{\circ}$. These impurity peaks can be indexed to the hexagonal, FM Mn$_5$Ge$_3$. The same impurity phase was present in samples of CaMn$_2$Ge$_2$ and BaMn$_2$Ge$_2$ that were previously investigated using neutron diffraction. \cite{Malaman-1994} A two-phase refinement, using SrMn$_2$Ge$_2$ and Mn$_5$Ge$_3$ as the phases, resulted in an improved fit to the data and resulted in a weight fraction of 5.1(2)\% Mn$_5$Ge$_3$. 

The tetragonal lattice parameters $a$ and $c$, as well as the  variable $c$-axis Ge position parameter $z_{\rm Ge}$ obtained from the two-phase refinement are listed in Table~\ref{Table:Structure} together with the corresponding parameters reported for CaMn$_2$Ge$_2$, BaMn$_2$Ge$_2$, and the relevant rare-earth  counterparts LaMn$_2$Ge$_2$ and EuMn$_2$Ge$_2$. In addition, the intralayer Mn-Mn distance $d^{ab}_{\rm Mn-Mn}$, as well as the interlayer Ge-Ge distance $d_{\rm Ge-Ge}$ [Fig.~\ref{Fig:XRD_Ge}] estimated from the refined/reported lattice parameters are also included in Table~\ref{Table:Structure}. The relatively small $c/a$ ratio, together with $d_{\rm Ge-Ge}$ values close to the single covalent Ge--Ge bond distance of $\sim 2.45$~\AA\, suggest the presence of interlayer bonding between Ge atoms, leading to a cT structure. It has been observed that interlayer bonding between metalloids has a direct influence on the ionic state of the transition-metal ions and, consequently, on the magnetic properties of the compounds.\cite{Pandey-2013b} This will be further discussed in the following sections.  

\subsection{Magnetic properties}
\label{Sec:Ge_Mag}

\begin{figure}
\includegraphics[width=3.1in]{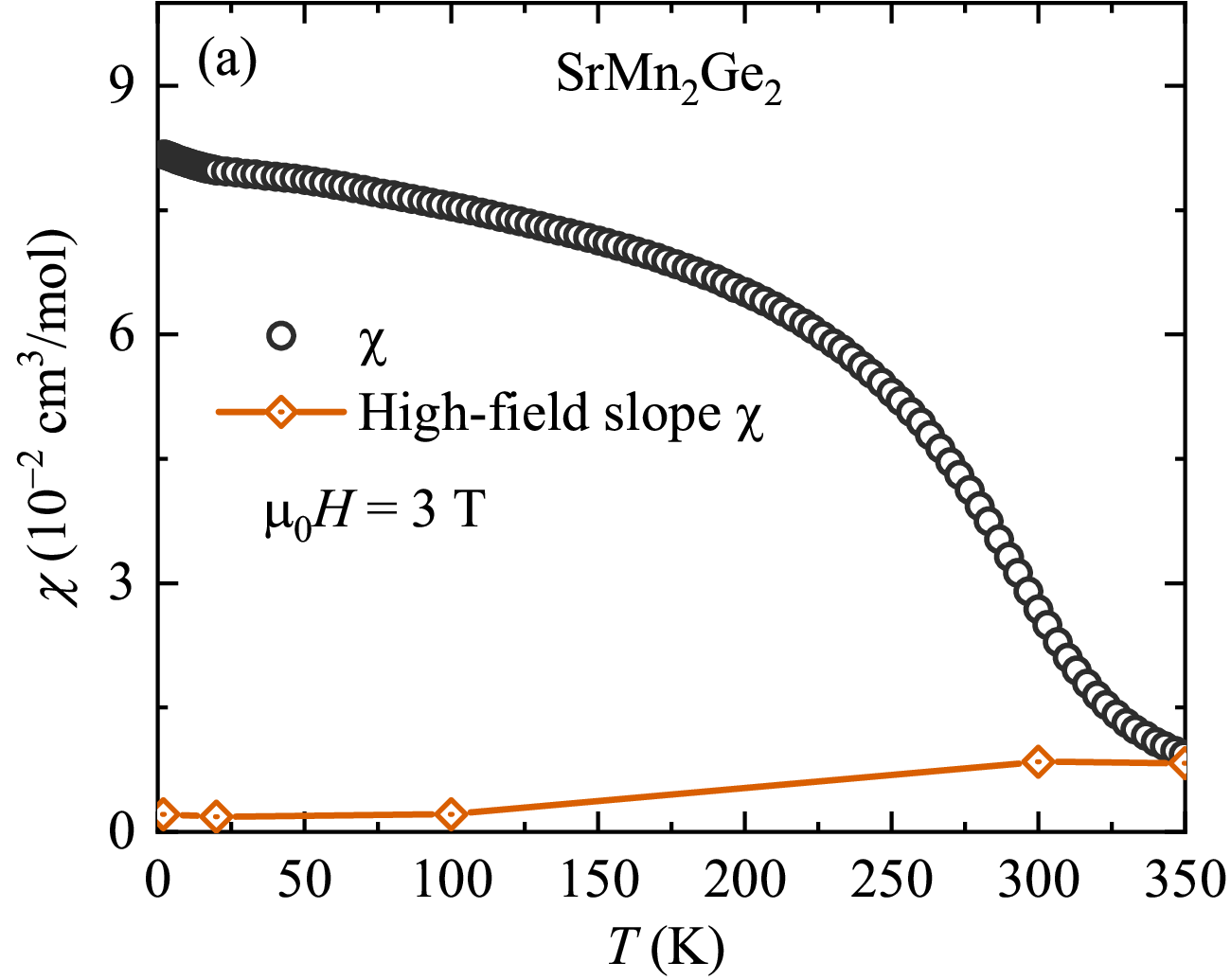}
\includegraphics[width=3.1in]{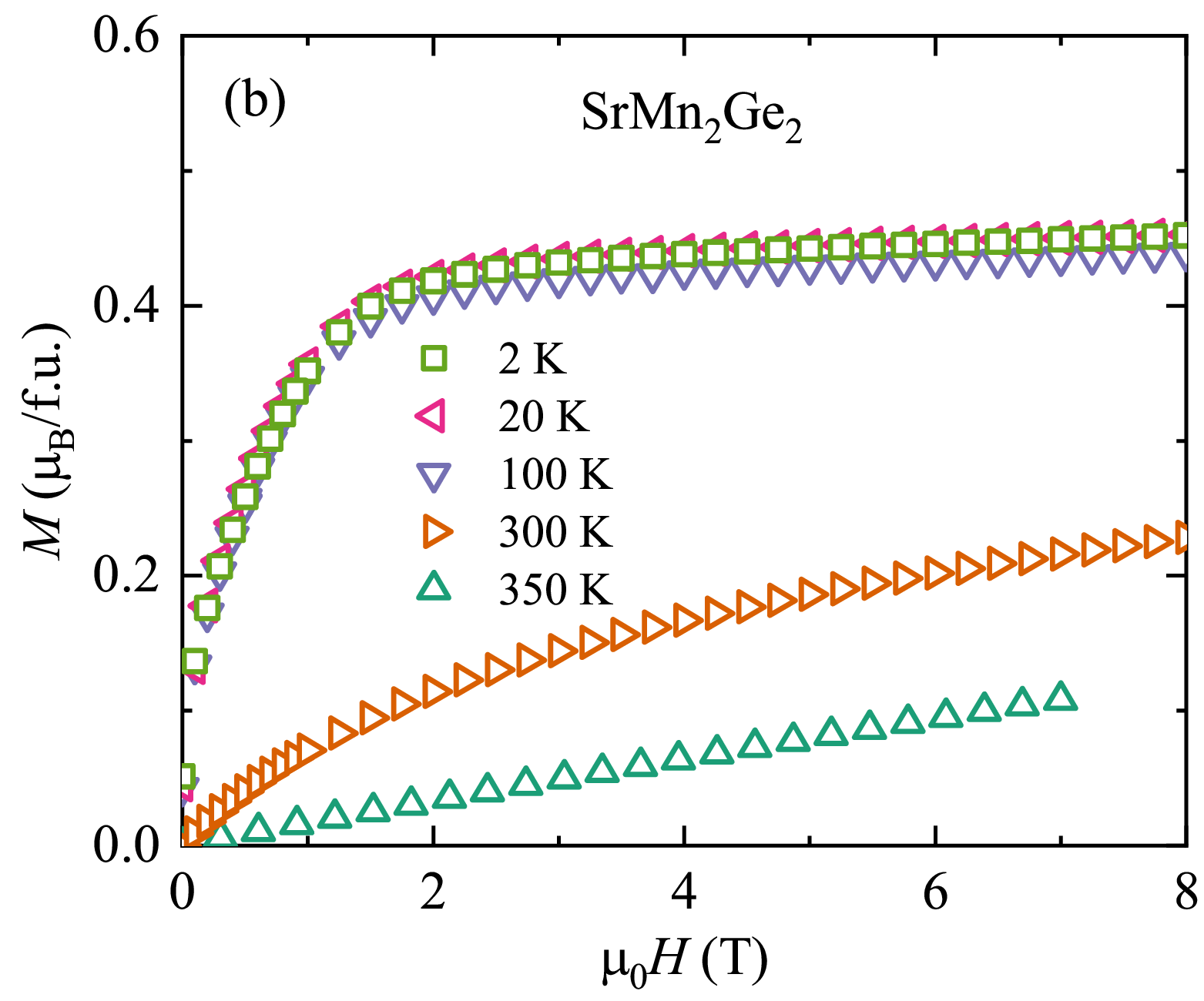}
\caption{(a) Zero-ﬁeld-cooled magnetic susceptibility $\chi = $ $M$/$H$ of SrMn$_2$Ge$_2$ as a function of  temperature $T$, measured in an applied magnetic ﬁeld $\mu_{\rm 0} H$ = 3~T.  The orange diamonds represent the $\chi$ values obtained from linear fits to the magnetic isotherms $M(H)$ shown in panel (b) for $\mu_0H >2$~T. The solid orange line connecting the diamonds is a guide to the eye. (b) Magnetization $M$ versus $H$ at five different temperatures.}
\label{Fig:Chi_Ge}
\end{figure}

The zero-field-cooled (ZFC) magnetic susceptibility $\chi = $ $M$/$H$ data of SrMn$_2$Ge$_2$ show a large increase in $\chi$ as temperature $T$ is decreased below $350$~K [Fig.~\ref{Fig:Chi_Ge}~(a)]. While the observed $T$-dependence of $\chi$ mimics that of an FM system, its small value at $T = 2$~K, $\chi_{\rm 2~K} = 8.2\times10^{-2}$~cm$^3$/mol, does not support the existence of an FM ground state in this compound. The same conclusion can be drawn from the low-$T$ ordered moment $M_{\rm o}$ deduced from the isothermal magnetization $M(H)$ data discussed below. Furthermore, the observed large increase in $\chi$ is consistent with the presence of small amounts of an FM impurity (with $T_{\rm C}\approx 300$~K) embedded within an AFM or non magnetic system. In such a case, a substantial magnetic contribution of the FM impurity below its Curie temperature $T_{\rm C}$ can mask the intrinsic low-$\chi$ or -$M$ of the compound. A similar observation was previously made in the case of a related compound BaMn$_2$As$_2$. \cite{Singh 2009} As discussed in the previous section, the two-phase refinement performed on the XRD data reveals the presence of about 5 wt.\% of FM impurity Mn$_5$Ge$_3$. Depending upon the sample, the $T_{\rm C}$ of Mn$_5$Ge$_3$ is reported to vary between 290 and 305~K\@. \cite{Songlin 2002, Synoradzki-2022} This suggests that the increase in $\chi$ with decreasing $T$ could likely be originating from the contribution of the embedded impurity phase and not an intrinsic property of SrMn$_2$Ge$_2$.\\

In order to determine the intrinsic susceptibility of SrMn$_2$Ge$_2$, corrected for the presence of FM impurities, we now examine the isothermal $M(H)$ data measured at temperatures between 2 and 350~K [Fig.~\ref{Fig:Chi_Ge}(b)]. The $M(H)$ curves exhibit a negative curvature at low magnetic fields ($\mu_{0}H \lesssim 2$~T), which becomes more pronounced with decreasing temperature. As $H$ increases above $\sim 2$~T, the $M(H)$ isotherms tend toward saturation and subsequently display nearly linear or slightly sub-linear behaviour. The tendency to saturation and the resulting ordered moment can be attributed to Mn$_5$Ge$_3$ impurity, which has an average magnetic moment of 2.64~$\mu_{\rm B}$ per Mn atom. \cite{Liu 2006} A $5.1\%$ wt fraction of Mn$_5$Ge$_3$ will result in a measured saturation magnetization of $\sim 0.5$~$\mu_{\rm B}$, which is close to the $M_{\rm o}\approx 0.43~\mu_{\rm B}$ obtained at 2~K. To correct the $\chi(T)$ data for the presence of FM impurities, we determined the high-field slopes of the $M(H)$ isotherms by performing linear fits to the data for fields $\mu_0H > 3$~T. Since FM contributions typically saturate at high fields ($\mu_0H \gtrsim 1$~T), the high-field slopes provide the intrinsic susceptibility $\chi_{\rm int}$ of the sample under consideration, corrected for impurity contributions. The $\chi_{\rm int}$ values of SrMn$_2$Ge$_2$ obtained in this manner are plotted as orange diamonds in Fig.~\ref{Fig:Chi_Ge}~(a). We find that $\chi_{\rm int}<\chi$ throughout the measured temperature range and $\chi_{\rm int}<< \chi$ for $T\leq 275$~K where the FM impurity begins to saturate. Furthermore, $\chi_{\rm int}$ exhibits a decrease with decreasing temperature which is consistent with the hypothesis that SrMn$_2$Ge$_2$ is an AFM with $T_{\rm N}>350$~K, as is reported for the neighbouring compounds CaMn$_2$Ge$_2$ and BaMn$_2$Ge$_2$, as well as related compound EuMn$_2$Ge$_2$, which has similar lattice parameters and G-type AFM structure (Table~\ref{Table:Structure}). \cite{Malaman-1994} The value of $\chi_{\rm int} = 2.1 \times 10^{-3}$~cm$^3$/mol at $T = 2$~K is similar to what is reported for the related compounds BaMn$_2$$Pn_2$ ($Pn = $ P, As, Sb, Bi). \cite{Singh 2009, Jacobs-2023, Sangeetha-2018, Saparovn-2013} While temperature-dependent neutron diffraction measurements are required to confirm the exact nature of the magnetic ground state, this agreement further supports the consistency of the above approach and suggests for a long-range AFM ground state for SrMn$_2$Ge$_2$.

\subsection{Electrical transport and magnetoresistance}

\begin{figure*}
\includegraphics[width=3.1in]{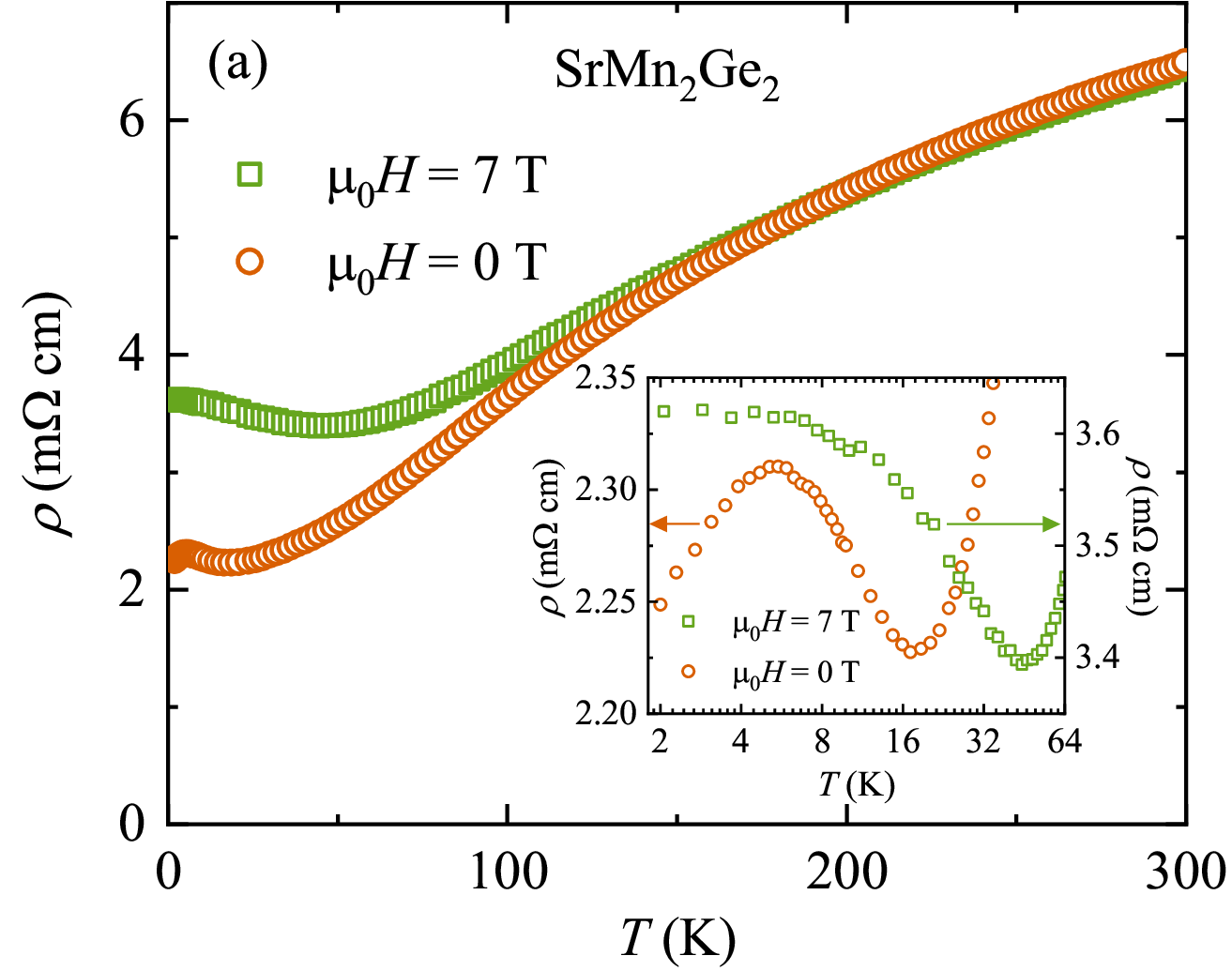}
\includegraphics[width=3.1in]{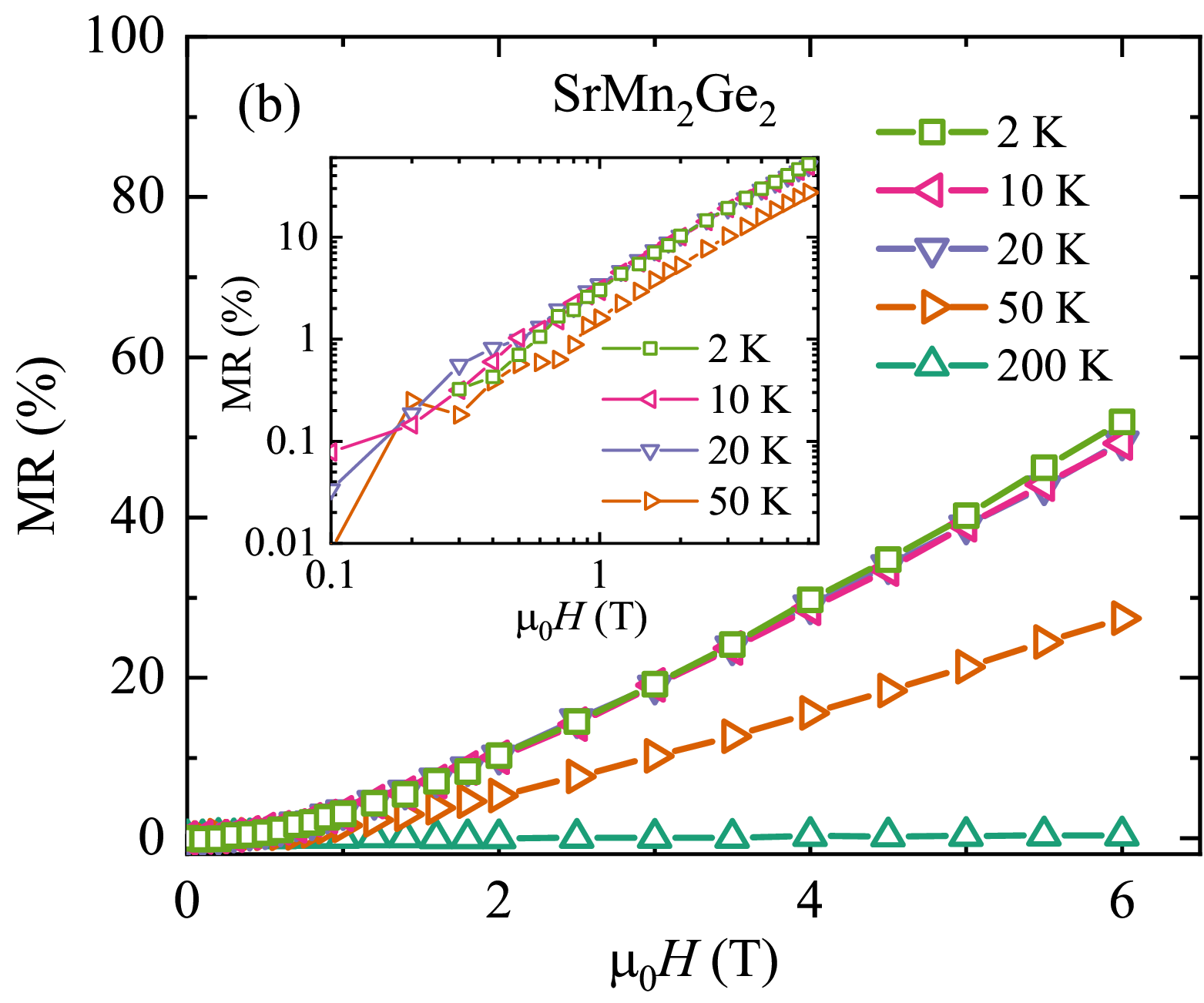}
\caption{(a) Resistivity $\rho$ of SrMn$_2$Ge$_2$ as a function of temperature $T$ at two different fields $\mu_0H = 0$ and 7~T\@. Inset: Low-$T$ features of $\rho(T)$ displayed on a semi-log scale. (b) Magnetoresistance (MR) versus $H$ at five different temperatures. Inset: MR versus $H$ for $T\leq 50$~K on a log-log scale.}
\label{Fig:Res_Sr}
\end{figure*}

Electrical resistivity $\rho$ of SrMn$_2$Ge$_2$ exhibits a monotonic increase with increasing $T$ at high temperatures and has a finite value at the lowest $T$ of our measurement, highlighting the metallic nature of the compound [Fig.~\ref{Fig:Res_Sr}~(a)]. The observed metallic behavior is consistent with heat capacity measurements and electronic structure calculations discussed later in the text. Furthermore, at low temperatures and in the absence of an applied field, the $\rho(T)$ data exhibits Kondo-like features, where a resistivity minimum and maximum appear at $T\sim 17$ and $\sim 5.5$~K, respectively. These features are highlighted on a semi-log scale in the inset of Fig.~\ref{Fig:Res_Sr}~(a). The high-field $\rho(T)$ data measured at $\mu_0H = 7$~T overlaps with the zero-field data for $T\gtrsim$~160~K, below which large and positive MR is observed. Such large and positive MR is unusual but not unique, and has been previously reported in the related compound LaMn$_2$Ge$_2$, where MR values approaching 100\% were observed at 4.2~K under an applied field of 7~T. \cite{Mallik 1997} The Kondo-like features observed at low temperatures in the absence of an applied field are smoothed out and partially suppressed upon application of the high field, resulting in a nearly constant $\rho$ below $T\sim 7$~K [Inset, Fig.~\ref{Fig:Res_Sr}~(a)]. Additionally, the resistivity minimum located near $T\sim 17$~K in zero field broadens and shifts to $T\sim 44$~K under $\mu_0H= 7$~T suggesting a magnetic origin. 

Fig.~\ref{Fig:Res_Sr}(b) shows the MR~$=\frac{\rho(H)-\rho(0)}{\rho(0)}$ versus $H$, measured at five temperatures between 2 and 200~K\@. The measured MR values are consistently positive and increase monotonically with increase in $H$. Additionally, the MR of SrMn$_2$Ge$_2$ does not follow a quadratic field dependence, which indicates the presence of correlated multiband electrical transport and/or complex magnetic scatterings. When plotted on a log-log scale, the MR data for $T \leq 50$~K exhibit the same sub-quadratic power-law dependence for all isotherms [Inset, Fig.~\ref{Fig:Res_Sr}~(b)]. Linear fits to these MR isotherms for fields $\geq 1$~T results in a power-law exponent of $n = 1.50(8)$.

\subsection{Heat capacity}
\label{Sec:HC_Ge}

\begin{figure}
\centering
\includegraphics[width=3.1in]{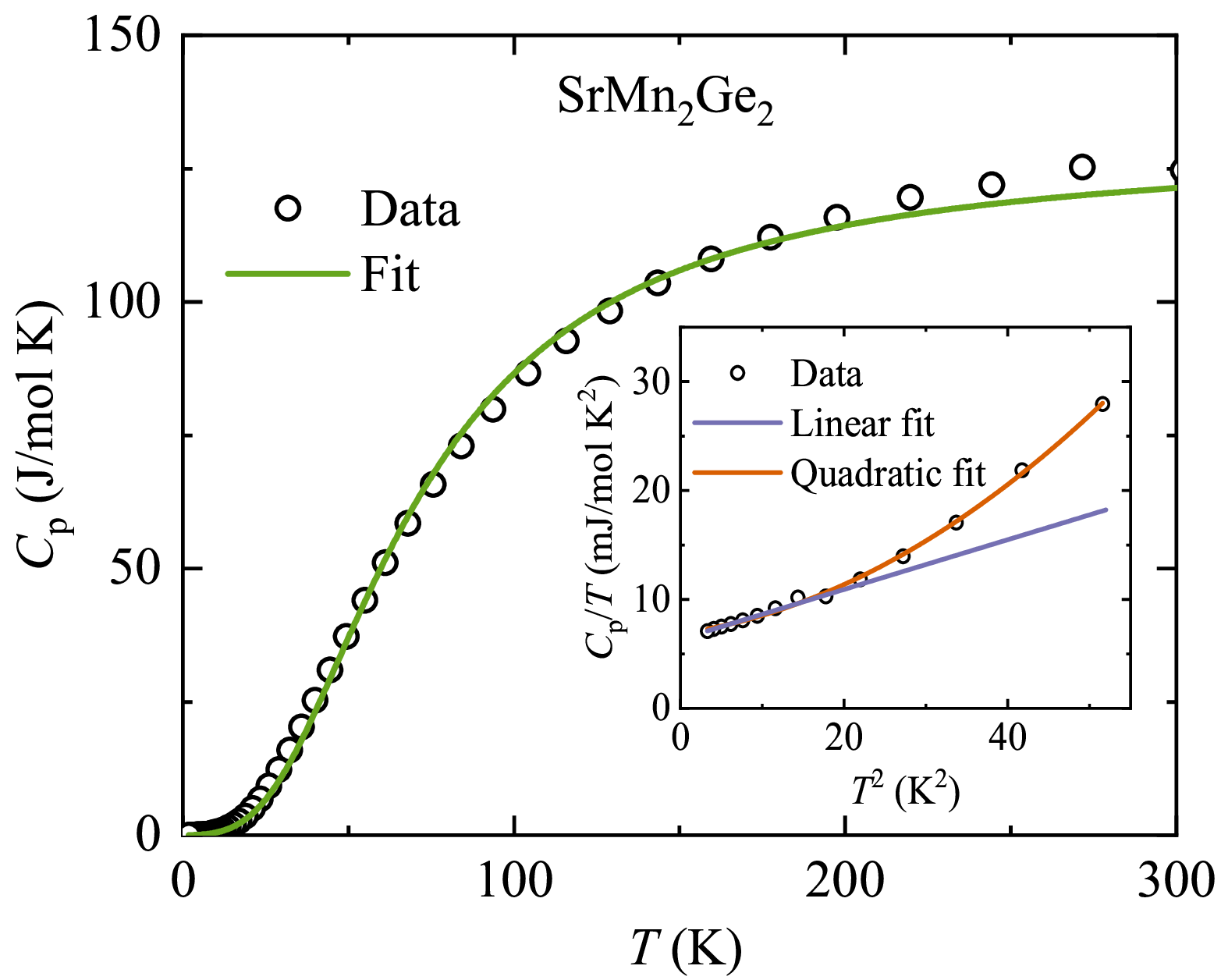}
\caption{Heat capacity $C_{\rm p}$ as a function of temperature  $T$ for SrMn$_2$Ge$_2$. The solid green curve is a Debye model fit to the data as described in the text. Inset: $C_{\rm p}/T$ vs $T^2$ at low temperatures. The solid curve corresponds to a fit to the data using Eq.~(\ref{Eq:HC_LowTemp2}a), while the solid line represents a fit using Eq.~(\ref{Eq:HC_LowTemp2}b).}
\label{Fig:HC_Ge}
\end{figure}

The isobaric heat capacity $C_{\rm p}(T)$ of SrMn$_2$Ge$_2$ does not show any evidence of a phase transition within the measured range of 1.8~K~$\le T \le$~300~K [Fig.~\ref{Fig:HC_Ge}]. Furthermore, the $C_{\rm p}$ data saturate to the Dulong-Petit limit of $C_{\rm V} = 3nR = 15R = 124.7$~J/(mol~K) at room temperature, where $n = 5$ is the number of atoms per formula unit and $R$ is the molar gas constant. 

The $C_{\rm p}/T$ versus $T^2$ data exhibits a linear behavior for  $T < 4$~K, but over all exhibits a positive curvature at low temperatures [Inset, Fig.~\ref{Fig:HC_Ge}], indicating the possible dominance of a higher-order term proportional to $T^5$ in lattice contribution to the $C_{\rm p}$. A similar positive curvature is observed in the related Mn-based insulator BaMn$_2$P$_2$. \cite{Jacobs-2023} Thus, the low-$T$ $C_{\rm p}/T$ data were fitted using two different equations: Eq.~(\ref{Eq:HC_LowTemp2}a), which indicates the dominance of a higher-order term manifested as a positive curvature, and Eq.~(\ref{Eq:HC_LowTemp2}b), which assumes its absence, leading to a linear variation below 4~K,
\begin{subequations}
\begin{align}
C_{\rm p}/T =  \gamma + \beta T^2+ \delta T^4,\\
C_{\rm p}/T =  \gamma + \beta T^2,
\end{align}
\label{Eq:HC_LowTemp2}
\end{subequations}
where $\beta$ and $\delta$ are coefficients describing the lattice vibrational contribution to $C_{\rm p}$ at low temperatures, and $\gamma$ is the Sommerfeld coefficient representing the electronic contribution to $C_{\rm p}$. A good fit to the data was obtained using both methods, resulting in the following fitted parameters; Eq.~(\ref{Eq:HC_LowTemp2}a): $\gamma = 6.9(2)$~mJ/mol~K$^2$, $\beta = 0.11(2)$~mJ/mol~K$^4$, and $\delta = 0.006(3)$~mJ/mol~K$^6$; and Eq.~(\ref{Eq:HC_LowTemp2}b): $\gamma = 6.3(3)$~mJ/mol~K$^2$ and $\beta = 0.24(2$~mJ/mol~K$^4$. The non-zero $\gamma$ confirms a metallic ground state of SrMn$_2$Ge$_2$, in agreement with the $\rho(T)$ data, as well as the electronic structure results described below. Moreover, the $\gamma$ can be used to estimate the experimental DOS at the Fermi level $E_{\rm F}$, $D^\gamma(E_{\rm F})$ using,
\begin{equation}
\gamma= \frac{\pi^2k_{\rm B}^2}{3}D^\gamma(E_{\rm F}).
\end{equation}
This yields $D^\gamma(E_{\rm F}) = 2.93(8)$~states/eV f.u [Eq.~(\ref{Eq:HC_LowTemp2}a)] and 2.7(2)~states/eV f.u [Eq.~(\ref{Eq:HC_LowTemp2}b)]. Furthermore, the Debye temperature $\Theta_{\rm D}$ can be calculated from the $\beta$ obtained above using,
\begin{equation}
\Theta_{\rm D}=\left( \frac{12 \pi^4 R n}{5 \beta}\right)^{1/3}.
\end{equation}
The estimated values are $\Theta_{\rm D} = 445(20)$~K [Eq.~(\ref{Eq:HC_LowTemp2}a)] and 346(4)~K [Eq.~(\ref{Eq:HC_LowTemp2}b)]. The $\Theta_{\rm D}$ obtained from Eq.~(\ref{Eq:HC_LowTemp2}a) is significantly larger than those reported for BaMn$_2$P$_2$ and BaMn$_2$As$_2$, namely 348(17) and 246(4)~K, respectively.\cite{Singh 2009,Jacobs-2023} In contrast, the $\Theta_{\rm D}$ estimated from Eq.~(\ref{Eq:HC_LowTemp2}b) is in better agreement with the reported values. Hence, it remains unclear if the use of a higher-order term that seems to dominant at elevated temperatures is physically realistic or not. Such behaviors may also indicate the presence of some sort of structural or electronic anomaly influencing the low-temperature properties.   

The $C_{\rm p}(T)$ data over the measured temperature range of 1.8~K~$<T<300$~K were fitted using 
\begin{equation}
C_{\rm p}(T) = \gamma T + nC_{\rm{V\,D}}(T),
\label{eq:HC}
\end{equation}
where $n = 5$ and $\gamma = 6.9(2)$~mJ/mol~K$^2$ were held fixed to the value obtained above. Here $C_{\rm{V\,D}}(T)$ is the Debye model heat capacity of acoustic phonons given by the expression 
\begin{equation}
    C_{\rm{V\,D}}(T)=9R\left(\frac{T}{\Theta_{\rm D}}\right)^3\int^{\Theta_D / T}_0 \frac{x^4e^x}{(e^x-1)^2}dx,
    \label{Eq:HC_Debye}
\end{equation}
A high-accuracy Pad\'{e} approximant\cite{Goetsch 2012} of the integral in Eq.~\ref{Eq:HC_Debye} was used, and the obtained fit was satisfactory [Fig.~\ref{Fig:HC_Ge}]. The fitted $\Theta_{\rm D}=284(3)$~K, which is significantly lower than that obtained from the low-temperature $C_{\rm p}(T)$ data, appears more reasonable, as the $C_{\rm p}(T)$ data show a clear tendency to saturate toward the Dulong-Petit limit near room temperature. While the exact reason for this descripnacy is not clear, we anticipate that this might arise because of the $T$-dependence of $\Theta_{\rm D}$, which is observed in several materials.\cite{Gopal}

\subsection{Electronic structure} 

ThCr$_2$Si$_2$-type compounds with high $c/a$ ratios and ucT tetragonal structure often exhibit quasi-2D features in their electronic and magnetic properties. Additionally, the electronic band structures of these compounds often feature flat bands which, if in close proximity to the Fermi level $E_{\rm F}$, can result in substantial electronic instabilities. \cite{Pandey-2013b} Given the relatively small $c/a$ ratio exhibited by SrMn$_2$Ge$_2$, it was unclear whether quasi-2D properties and flat bands will occur in this compound. To investigate the potential occurrence of these features in the electronic structure of SrMn$_2$Ge$_2$, we first computed the total energy for the FM configuration, as well as  three different AFM configurations---A-type, C-type, and G-type (Table~\ref{Table:TotalEnergy}). The G-type AFM configuration, adopted by CaMn$_2$Ge$_2$ and BaMn$_2$Ge$_2$,\cite{Malaman-1994} resulted in the lowest total energy. Using this AFM configuration, we obtain magnetic moments of 3.28 and 3.77~$\mu_{\rm B}$/Mn atom with PBEsol and r$^2$SCAN, respectively.

\begin{figure}
\centering
\includegraphics[width=3.4in]{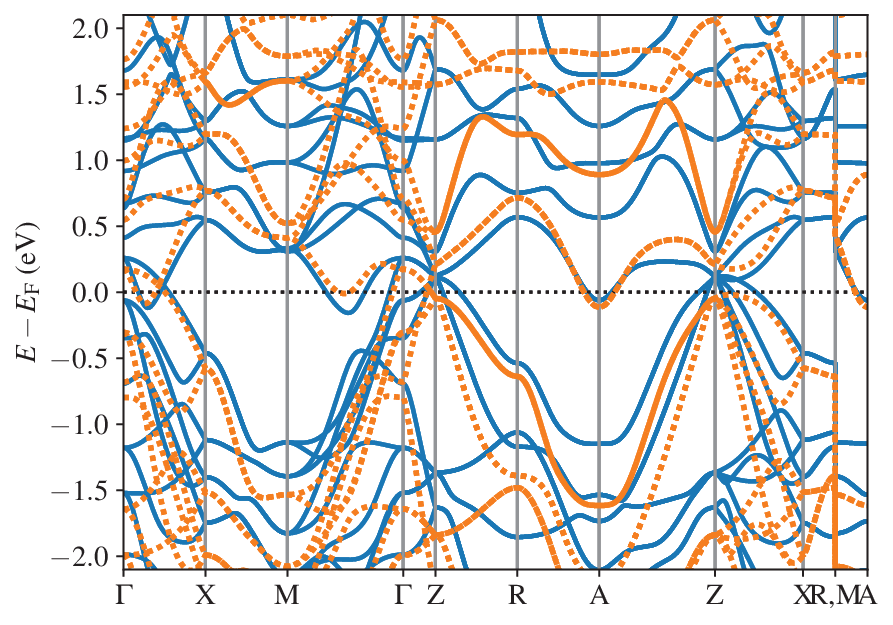}
\caption{Electronic band structure of
SrMn$_2$Ge$_2$ calculated using PBEsol (solid blue lines) and r$^2$SCAN (dotted orange lines). All calculations have been performed using the G-type AFM configuration. }
\label{Fig:Bands_comparison_Ge}
\end{figure}

\begin{figure*}
\centering
\includegraphics[width=4.2in]{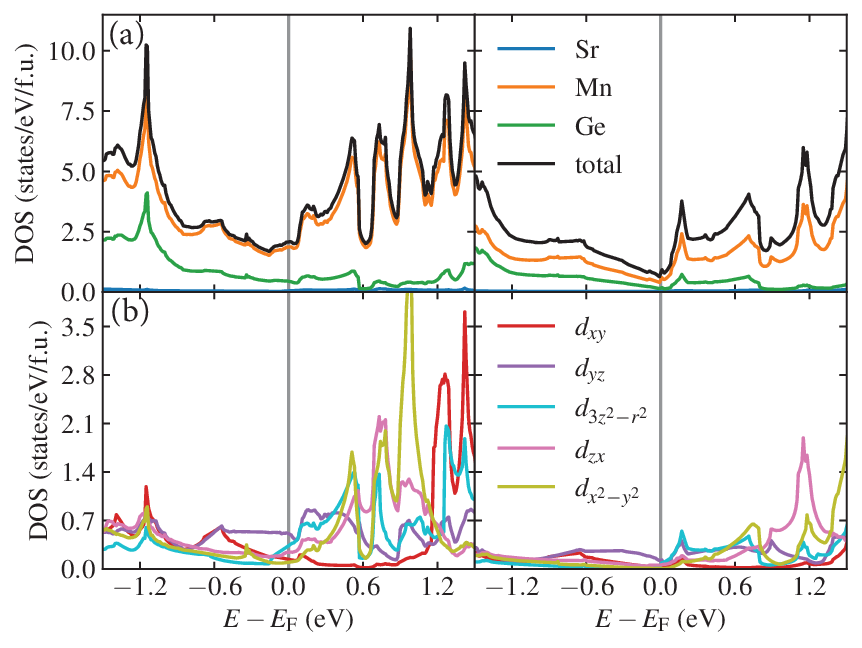}
\caption{(a) Total and atom-decomposed density of States (DOS) computed with PBEsol (left panel) and r\textsuperscript{2}SCAN (right panel), and (b) orbital-decomposed DOS computed with PBEsol (left panel) and r\textsuperscript{2}SCAN (right panel) for SrMn$_2$Ge$_2$ using the G-type AFM configuration.}
\label{Fig:DOS_Ge}
\end{figure*}

The electronic band structure of SrMn$_2$Ge$_2$ features a confluence of bands at the $Z$-point, with multiple crossings just above the $E_{\rm F}$ [Fig.~\ref{Fig:Bands_comparison_Ge}]. The r$^2$SCAN, which typically captures correlation effects to a greater extent, pushes these bands apart reducing the metallicity of the compound. This behavior is further evidenced by Fig.~\ref{Fig:DOS_Ge}, where the $D^{\rm band}(E_{\rm F})$ is found to decrease when computed using the r$^2$SCAN functional resulting in 0.8 states/(eV~f.u.) compared to the PBEsol value of 2.1 states/(eV~f.u.). These values are lower than the $D^\gamma(E_{\rm F})$ we obtained from the $C_{\rm p}(T)$ data (see Section~\ref{Sec:HC_Ge}) indicating the presence of substantial electronic correlation effects and associated quasiparticle mass enhancement in SrMn$_2$Ge$_2$. Further, the band structure of SrMn$_2$Ge$_2$ does not exhibit dispersion-less bands near $E_{\rm F}$, suggesting the absence of electronic instabilities in this compounds. The closest flat band to the $E_{\rm F}$ is located approximately 0.75~eV above $E_{\rm F}$ between the $X$ and $R$ points. Hence, we conclude that the electronic structure exhibits 3D characteristics, which is consistent with the  crystal structure revealing a relatively low $c/a$ ratio and a short $d_{\rm Ge-Ge}$.

\begin{table} 
\centering 
\caption{Density functional theory total energies per formula unit in electron volts for the different magnetic configurations, relative to the G-type antiferromagnetic (AFM-G) ground state, calculated using the PBEsol and r$^{2}$SCAN exchange-correlation functionals. The abbreviations are defined as follows: nonmagnetic (NM); ferromagnetic (FM); and antiferromagnetic (AFM).}
\label{Table:TotalEnergy}
\hspace*{-0.4cm}
\begin{tabular}{l l l }
\hline
     Ordering type & PBEsol & r$^{2}$SCAN  \\
	 \hline
      NM & 1.742 & 3.797 \\
			FM & 0.403 & 0.384 \\
			AFM-A & 0.382 & 0.477 \\
			AFM-C &  0.142 & 0.145 \\
			AFM-S & 0.366 & 0.358 \\
			AFM-G & 0 &  0 \\
  \hline
            
\end{tabular}
\end{table}

\section{Discussion}

To gain deeper insight, the structural and physical properties of tetragonal compounds AMn$_2$Ge$_2$ ($A =$ Ca, Sr, Ba) need to be evaluated in comparison to those of LaMn$_2$Ge$_2$ and BaMn$_2$$Pn_2$ ($Pn =$ P, As, Sb, Bi). On one hand, LaMn$_2$Ge$_2$ appears structurally closer to SrMn$_2$Ge$_2$, exhibiting a similar $c/a$ ratio and $d_{\rm Ge-Ge}$. On the other hand, BaMn$_2$$Pn_2$ shows similarities of magnetic ground states, as infered by the same G-type AFM spin structure and similar values of observed Mn magnetic moment. \cite{Malaman-1994,Johnston-2011}

It has been established that in ThCr$_2$Si$_2$-type 122 pnictides, the interlayer coupling between the $Pn$-ions controls $d_{\rm Pn-Pn}$. Consequently, $d_{\rm Pn-Pn}$ correlates with the effective ionic state of the transition metal ions, and as a result, with the magnetic ground state of the compounds.\cite{Pandey-2013b} The uncollapsed tetragonal (ucT) structure of BaMn$_2$$Pn_2$ infers a weak interlayer coupling between the $Pn$-ions, leading to a $Pn^{-3}$ valence state. This further results in a ${+2}$ valence state of the Mn ions. In contrast, for collapsed tetragonal (cT) SrMn$_2$Ge$_2$, the small $d_{\rm Ge-Ge}$ indicates interlayer Ge-Ge dimerization. Following the analogy of the $Pn$-counterparts, this dimerization would result into the Ge valence valency that is likely similar to those of $Pn$-ions. This could stabilize a ${+2}$ valence state of the Mn ions in SrMn$_2$Ge$_2$, and hence, result in a magnetic moment similar to those of BaMn$_2$$Pn_2$. We emphasize that the ionic valence picture often breaks down in metallic systems. Therefore, the above discussion is intended only to provide a qualitative understanding of the similar magnetic features of SrMn$_2$Ge$_2$ and BaMn$_2$$Pn_2$. Neutron diffraction measurements are essential to explore the underlying magnetic structure of the material. Additionally, detailed spectroscopic and transport properties studies are required to achieve a deeper understanding of the complex mechanisms that govern the magnetism and its manifestations in this unexplored compound.

Furthermore, the magnetic interactions and the resulting properties of ucT semiconducting BaMn$_2$$Pn_2$ compounds have been successfully described using $J_1$-$J_2$-$J_c$ Heisenberg model of stacked-square lattice of Mn-moments.\cite{Johnston-2011}. Here, $J_{1}$ and $J_{2}$ represent the nearest and next-nearest neighbor intralayer exchange interactions within the Mn square lattice, respectively, while $J_{c}$ denotes the interlayer exchange interaction between adjacent Mn layers along the crystallographic $c$-axis. It would be interesting the explore whether the same model remains applicable in the case of cT metallic SrMn$_2$Ge$_2$, where metallic nature may lead to a pronounced itinerant character and the three-dimensional character of the underlying magnetic sublattice may require additional exchange parameters.  

The observed magnetoresistance (MR) of SrMn$_2$Ge$_2$ is intriguing and indicates that a multiband conduction mechanism, as well as complex magnetic scattering processes are likely present and dominant. Here, of course, we assume that the presence of small amount of Mn$_5$Ge$_3$ impurity does not significantly affect the conduction mechanism within the primary phase. Somewhat similar observations of positive MR have been reported in LaMn$_2$Ge$_2$.\cite{Mallik 1997} However, in contrast to a power-law dependence of MR on $H$ in SrMn$_2$Ge$_2$, the MR in LaMn$_2$Ge$_2$ exhibits a nearly linear dependence on $H$. Additionally, LaMn$_2$Ge$_2$ displays complex competing magnetic states below its $T_{c} = 326$~K, which can be described as an evolving noncollinear conical spin structure with a ferromagnetic component.\cite{Gong-2021} Therefore, a direct comparison of the magnetotransport properties of SrMn$_2$Ge$_2$ vis-a-vis LaMn$_2$Ge$_2$ may not be prudent. Nevertheless, the intriguing observation of giant room-temperature topological Hall effect in LaMn$_2$Ge$_2$, which is relevant to spintronics applications, motivates a careful investigation of the magnetic ground state, as well as of the electronic transport mechanism in SrMn$_2$Ge$_2$, preferably using single-crystal samples.  

\section{Conclusion}
\label{Sec:Ge_Summary}

A comprehensive investigation of the physical properties of the previously unexplored tetragonal compound SrMn$_2$Ge$_2$ is reported. Powder x-ray diffraction data confirm the ThCr$_2$Si$_2$-type structure. Magnetic measurements are consistent with a possible antiferromagnetic ground state with a $T_{\rm N}> 350$~K\@; however, high-temperature measurements and neutron diffraction studies are necessary to confirm the magnetic ground state. The heat capacity data, together with electronic structure calculations, suggest for the presence of sizable electronic correlations in the material. Notably, SrMn$_2$Ge$_2$ exhibits an unusually large magnetoresistance of approximately 50\% at low temperatures, obeying a power-law field dependence with an exponent of $n = 1.50(8)$. This intriguing observation indicates the dominance of a multiband conduction mechanism, as well as the presence of complex magnetic scattering processes. Furthermore, the low-$T$ resistivity data exhibits Kondo-like features. The intriguing properties of SrMn$_2$Ge$_2$ warrant further investigation, particularly on high-quality single crystals, using microscopic probes such as neutron diffraction and scattering, NMR, ARPES, and XPS. Such studies would provide deeper insight into its magnetic and electronic properties and help understand the mechanisms that govern the observed phenomena.

\section*{Conflicts of interest}
There are no conflicts to declare.

\section*{Data availability}

Electronic structure input and output files are available at the open data repository with the following DOI: \url{https://doi.org/10.5281/zenodo.19330078}.\\ 

All data supporting the findings of this study are available within the article and/or its supplementary information. 

\section*{Acknowledgements}

This work is based on the research supported in part by the National Research Foundation of South Africa (Ref Numbers: SRUG2205098188). A.P. acknowledges the financial support from the Friedel Sellschop fellowship of the University of the Witwatersrand, South Africa. RW acknowledges the Centre for High Performance Computing (CHPC), South Africa, for providing computational resources to this research project.\\



\balance


\end{document}